\documentstyle[12pt]{article}

        \def\l{\label}
         \def\d{\dagger}
         \def\be{\begin{equation}}
         \def\bea{\begin{eqnarray}}
          \def\la{\lambda}
         \def\o{\over}
         \def\b{\beta}
         \def\ro{\rho}
         \def\a{\alpha}

         \def\ee{\end{equation}}
         \def\eea{\end{eqnarray}}
         \def\R{\rm {I\kern-.200em R}}
         \def\C{\rm {I\kern-.520em C}}
         \def\r{\ref}
         \def\c{\cite}
\begin{document}
\begin{titlepage}
\vspace*{5mm}
\begin{center} {\Large \bf Generalized simplicial chiral models }\\
\vskip 1cm
Masoud Alimohammadi \footnote {e-mail:alimohmd@theory.ipm.ac.ir} \\

\vskip 1cm
{\it  Physics Department, University of Teheran, North Karegar Ave.,} \\
{\it Tehran, Iran }\\
{\it  Institute for Studies in Theoretical Physics and Mathematics,}\\
{\it  P.O.Box 5531, Tehran 19395, Iran}
\end{center}
\vskip 2cm
\begin{abstract}
Using the auxiliary field representation of the simplicial chiral models on a
$(d-1)$--dimensional simplex, the simplicial chiral models are generalized
through replacing the term Tr$(AA^{\d})$ in the Lagrangian of these models
by an arbitrary class function of $AA^{\d}$; $V(AA^{\d})$.
This is the same method used in defining
the generalized two--dimensional Yang--Mills theories (gYM$_2$) from ordinary
YM$_2$. We call these models, the ``generalized simplicial chiral models''.
Using the results of the one--link integral over a U($N$) matrix,
the large--$N$ saddle--point equations for eigenvalue density function $\ro (z)$
in the weak ($\b >\b_c$) and strong ($\b <\b_c$) regions are computed. In $d=2$,
where the model is in some sense related to the gYM$_2$ theory, the
saddle--point equations are solved for
$\ro (z)$ in the two regions, and the explicit value of critical
point $\b_c$ is calculated
for $V(B)$=Tr$B^n$ $(B=AA^{\d})$. For $V(B$)=Tr$B^2$,Tr$B^3$, and Tr$B^4$,
the critical behaviour of the model at $d=2$ is studied,
and by calculating the internal energy, it is shown that
these models have a third order phase transition.

\end{abstract}

\begin{center}
\end{center}
\end{titlepage}
\newpage

\section{ Introduction }

One of the useful approaches in studying the large--scale structure of non--Abelian
gauge theories is via the lattice formulation introduced by Wilson \c{kw},
in which a matrix--valued field theory is defined on a lattice. After the work
of Wilson,
several other matrix models have been introduced and their properties have
been studied. One of the most important features of matrix models, is their
large--$N$ behaviour, where $N$ is a quantity related to the number of field
components. This goes back to the original work by Stanley \c{he} on the
large--$N$ limit of spin system with O($N$) symmetry, soon followed by Wilson's
suggestion that the $1/N$ expansion may be a valuable alternative in the context
of renormalization group evaluation of critical exponents, and by 't Hooft
extension \c{gt} to gauge theories and, more generally, to fields belonging
to the adjoint representation of SU($N$) groups. Therefore the $1/N$ expansion
is probably the most important nonperturbative and analytical
tool presently available in the study of matrix models. Unfortunately,
application of this tool
is limited to a small number of few--matrix
systems. This number is even smaller if we restrict ourselves to the case of
unitary matrix fields, which is especially relevant to the problem of lattice
QCD. To the best of our knowledge, the only solved examples are Gross--Witten's
single--link problem \c{dg} and its generalizations, the external field problem
\c{rc1,eb}, and $L=3,4$ chiral chains \c{rc2,df}.

One of the important classes of the unitary matrix models, as mentioned above, are
few unitary--matrix systems. The interest for few--matrix models may arise
for various reasons. For example, their large--$N$ solutions may represent
non--trivial bench--marks for new methods meant to investigate the large--$N$
limit of more complex matrix models, such as QCD. Furthermore, every matrix
system may have a role in the context of two--dimensional quantum gravity;
indeed, via the double scaling limit, its critical behaviour is connected to
two--dimensional models of matter coupled to gravity. For more discussion
about these models see ref. \c{pr1}.

An interesting class of finite--lattice matrix models is obtained by considering
the possibility that each of a finite number of unitary matrices may interact
in a fully symmetric way with each other, while preserving global
chiral invariance; the resulting system can be described as a chiral model on
a $(d-1)$--dimensional simplex. A simplex is formed by connecting $d$ vertices
by $d(d-1)/2$ links. These models are known as ``simplicial chiral models''
(SCM) \c{pr2}. In ref. \c{rc3}, the large--$N$ saddle--point equations for
density function $\ro (z)$ of these models have been found.
The main strategy for the determination of these
equations, is based on the introduction of a
single auxiliary variable $A$ (a complex matrix), leading to the decoupling
of the unitary matrix interaction. The resulting action, contains a
Tr$(AA^{\d})$ term and some other linear terms in $A$ and $A^{\d}$.
Performing the single--link external field integral, the authors of \c{pr2}
have found the saddle--point equations.
In $d=2$ (where
the model corresponds to the Gross--Witten single--link problem, which in turn
is equivalent to large--$N$ QCD$_2$ with Wilson action on the lattice), $d=4$,
and $d\rightarrow \infty$, the saddle--point equations have been studied
analytically. It was shown that the critical value of $\b (=(g^2N)^{-1}$, where
$g$ is coupling constant) is $\b_c=1/d$ for all $d$. Also for $0\leq d <4$,
it has been shown that the
model exhibits the third--order phase transition, and for $4<d<\infty$ it has a
first--order phase transition.

On the other hand, it is known that the pure 2--dimensional Yang--Mills theory
(YM$_2$), with Tr($F^2$) Lagrangian, can be also represented by the Lagrangian
$i$Tr($BF$)+ Tr($B^2$) , in which $B$ is an auxiliary pseudo--scalar field in
the adjoint representation of the gauge group. Path integration over $B$, leaves
an effective Lagrangian Tr($F^2$). Now the generalized
2--dimensional Yang--Mills theory (gYM$_2$) is a theory with the above Lagrangian
in which the term Tr($B^2$) is replaced by an arbitrary class function $f(B)$
\c{ew}. The partition function of gYM$_2$ has been calculated in different
contexts in \c{ew}, \c{mr}, and \c{mk}. Also the phase structure of these
models, at large--$N$ limit, have been studied in \c{br}, in \c{ma1} for
$f(B)$=Tr($B^4$), and in \c{ma2} for $f(B)$=Tr($B^6$) and Tr$(B^2)+g$Tr$(B^4)$.
In \c{ma1} and \c{ma2}, it has been shown that these models have a third--order
phase transition, as ordinary YM$_2$.

In this paper, we want to introduce and study the generalized simplicial
chiral models (gSCM), in the same fashion that gYM$_2$ has been defined, that
is, through replacement of the term Tr($AA^{\d}$) in SCM by $V(AA^{\d})$, where
$V(AA^{\d})$ is an arbitrary class function of $AA^{\d}$. In this way, we have
a new few--matrix model on a $(d-1)$--simplex which, in the special case
$V=$Tr$(AA^{\d})$,
reduces to ordinary SCM. In $d=2$, where this model is in some sense related
to gYM$_2$,
we see that the study of the phase structure of the model is much easier than
the corresponding studies in gYM$_2$.

The paper is organized as follows.
In section 2 we introduce the gSCM and, using the single--link integral
method, find the large--$N$ saddle--point equations for eigenvalue density
function $\ro (z)$ in the weak and strong regions. Since these
equations for arbitrary $V(B)$ $(B=AA^{\d})$ are complicated, they can not
be solved analytically
for arbitrary $d$. Note that this is also true for $V(B)=$Tr$B$, i.e. SCM, and
as mentioned earlier, for very few cases  one can compute the density function
$\ro (z)$ analytically. Here the situation is worse and therefore we restrict
ourselves to $d=2$. In section 3, the general case
$V(B)=$Tr$(B^n)$, with $n$ an arbitrary positive integer number, is considered
and an
expression for internal energy in terms of $\ro (z)$ (for arbitrary $d$), and
also the critical value $\b_c$ for $d=2$ are found. In section 4, considering
the leading terms of $\ro (z)$ (near the critical point) in weak and strong
regions (again in $d=2$), it is shown that for $V(B)=$Tr$(B^2)$, Tr$(B^3)$, and
Tr$(B^4)$ the model exhibits a third order phase transition. We believe that
this behaviour (the third order phase transition) is the same for all
$V(B)=$Tr$(B^n)$ models.

\section{ gSCM and their saddle--point equation }

If we assign a U($N$) matrix to each vertex of a $(d-1)$--dimensional simplex,
then the partition function of simplicial chiral models is defined by \c{pr2}
\be \l{1}
Z_d(\b ,N)=\int \prod_{i=1}^ddU_i{\rm exp} \{ N\b \sum_{i=1}^d\sum_{j=i+1}^d
{\rm Tr} (U_iU_j^{\d}+U_i^{\d}U_j)\},
\ee
where $dU_i$ is the normalized invariant Haar measure. The
$d$--matrix simplicial model has an underlying permutation symmetry instead of
the cyclic symmetry of the $d$--matrix chiral chain. But for $d=1,2,$ and $3$
these two symmetries and the associated models are equivalent. The free energy,
internal energy, and specific heat are respectively given by
$$ F_d(\b ,N)={1\o {N^2}}{\rm ln} Z_d(\b ,N),$$
$$ U_d(\b ,N)={1\o 2}{{\partial F_d(\b ,N)} \o {\partial \b}},$$
\be \l{2}
c_d(\b ,N)=\b^2{{\partial U_d(\b ,N)} \o {\partial \b}}.
\ee
As mentioned earlier, the main strategy for determination of the large--$N$
saddle--point equation is based on the introduction of a single auxiliary
variable $A$ (a complex matrix), leading to the decoupling of the unitary matrix
interaction \c{pr2}
\be \l{3}
Z_d=\tilde Z_d/\tilde Z_0,
\ee
where
\be \l{4}
\tilde Z_d=\int \prod_{i=1}^ddU_idA{\rm exp} \{-N\b{\rm Tr}AA^{\d}+N\b{\rm Tr}
A\sum_iU_i^{\d}+N\b{\rm Tr}A^{\d}\sum_iU_i-N^2\b d\}.
\ee
Performing the single--link integral over the matrices $U_i$
\be \l{5}
e^{NW(BB^{\d})}\equiv \int dU{\rm exp}[N{\rm Tr}(B^{\d}U+U^{\d}B)],
\ee
we obtain
\be \l{6}
\tilde Z_d=\int dA{\rm exp} \{-N\b{\rm Tr}AA^{\d}+NdW(\b^2AA^{\d})-N^2\b d\}.
\ee

Now we define the generalized simplicial chiral model (gSCM) through the
partition function
\be \l{7}
Z_{d,V}=\tilde Z_{d,V}/\tilde Z_{0,V},
\ee
where
\be \l{8}
\tilde Z_{d,V}=\int dA{\rm exp} \{-N\b V(AA^{\d})+NdW(\b^2AA^{\d})-N^2\b d\},
\ee
in which $V(AA^{\d})$ is an arbitrary class function of $AA^{\d}$, that is
\be \l{9}
V(GAA^{\d}G^{-1})=V(AA^{\d}), \ \ \ \ \ \forall G\in U(N).
\ee
The requirement (\r{9}) is satisfied for every polynomial
\be \l{10}
V(AA^{\d})=\sum_{n=1}a_n{\rm Tr}(AA^{\d})^n.
\ee

The crucial point in our analysis is that the integrand in eq.(\r{8}) is
a function of eigenvalues $x_i$ of the Hermitian semi-positive--definite matrix
$\b^2AA^{\d}$. Moreover, as $\tilde Z_{d,V}$ is invariant under
$A\rightarrow G^{\d}AG'$ (where $G$ and $G'$ are arbitrary unitary matrices),
one can perform exactly the ``angular'' integration and reduce the problem to that
of integration over $N$ variables \c{tr}. Therefore, performing the angular
integration, eq.(\r{8}) leads to
\be \l{11}
\tilde Z_{d,V}=\int d\mu (x_i){\rm exp} \{-N\b V(x_i/\b^2)+NdW(x_i)-N^2\b d\},
\ee
where
\be\l{12}
d\mu (x_i)=\prod_idx_i\prod_{i>j}(x_i-x_j)^2.
\ee
In the large--$N$ limit, in which we are interested, the free--energy function
$W(x_i)$,
resulting from a single--link integral over a U($N$) matrix, can be extracted
by solving the Schwinger--Dyson equations. It is written in a simple closed form
\c{rc1,eb}
\be\l{13}
W(x_i)=2\sum_i\sqrt{x_i+c}-{1\o {2N}}\sum_{ij}{\rm ln}(\sqrt{x_i+c}+\sqrt{x_j+c})
-Nc-{3\o 4}N.
\ee
We must distinguish two different phases, a weak--coupling regime where $c=0$
and
\be\l{14}
{1\o {2N}}\sum_i{1\o{\sqrt{x_i}}}\leq 1,
\ee
and a strong--coupling regime where $c$ is dynamically determined by the
condition
\be\l{15}
{1\o {2N}}\sum_i{1\o{\sqrt{x_i+c}}}=1.
\ee
Therefore, in the large--$N$ limit the partition function (\r{11}) reduces to
\be\l{16}
\tilde Z_{d,V}=\int \prod_{i=1}^Ndx_ie^{-S_{d,V}(\{x_i\})},
\ee
with the action
$$ S_{d,V}(\{x_i\})=-2\sum_{i>j}{\rm ln}|x_i-x_j|+N\b V(x_i/\b^2)+N^2\b d $$
\be\l{17}
-Nd\{2\sum_i\sqrt{x_i+c}-{1\o {2N}}\sum_{ij}{\rm ln}(\sqrt{x_i+c}+\sqrt{x_j+c})
-Nc-{3\o 4}N\},
\ee
and the saddle--point equation $ \partial S / \partial x_i=0$ becomes
(after multiplying it by $(1/N)\sqrt{x_i+c}$ )
\be\l{18}
\b \sqrt{x_i+c} {d\o {dx_i}}V(x_i/\b^2)-d={1\o {2N}}\sum_{j\neq i}{{(4-d)
\sqrt{x_i+c}+d\sqrt{x_j+c}}\o {x_i-x_j}},
\ee
with the condition $x_i\geq 0$. To study eq.(\r{18}), we use the standard
technique which is based on the
eigenvalue density function. It is however convenient first to introduce a
new variable $z_i$ with definition
\be\l{19}
z_i=\sqrt{x_i+c},
\ee
subject to the condition $0\leq \sqrt{c} \leq z_i$. If the eigenvalue variable
$x_i$ varies in the interval $[x_a,x_b]$, $0\leq x_a\leq x_b$, then the new
variable $z_i$ lies in interval $[a,b]$ where $a=\sqrt{c+x_a}, b=\sqrt{c+x_b}$,
and
\be\l{20}
0\leq \sqrt{c} \leq a\leq b.
\ee
In the weak--coupling regime, $c=0$, one expects in general $a=\sqrt{x_a}>0$,
and in the strong--coupling regime, one expects $x_a=0$ so that
$a=\sqrt{c}\neq 0$ \c{rc3}.

Now, using the large--$N$ eigenvalue density function $\ro (z)$, which vanishes
outside the interval $[a,b]$, the saddle--point equation (\r{18}) can be turned
into the following integral equation
\be\l{21}
z\sum_{n=1}{{na_n}\o \b^{2n-1}}(z^2-c)^{n-1}-d={1\o 2}{\cal P}\int^b_adz'\ro (z')
\left( {2\o {z-z'}}-{{d-2}\o {z+z'}}\right),
\ee
in which we use the expression (\r{10}) for $V$ and ${\cal P}$ indicates the
principal value of integral. In this equation, the parameters $a$
and $b$ must be determined dynamically. The normalization condition of $\ro (z)$
is
\be\l{22}
\int_a^b\ro (z')dz'=1.
\ee
In the weak coupling regime, $c_w=0$ and the condition (\r{14}) in the
large--$N$ limit becomes
\be\l{23}
\int_a^bdz'{{\ro (z')}\o {z'}}\leq 2.
\ee
In the strong coupling regime, where c is
\be\l{24}
c_s=a^2,
\ee
the condition (\r{15}) becomes
\be\l{25}
\int_a^bdz'{{\ro (z')}\o {z'}}= 2.
\ee
Eq. (\r{21}) has a somewhat unconventional form when compared to other integral
equations, and one must perform a few manipulations to obtain a more
familiar equation. To do so, we define the function $H(z)$ in the complex--$z$
plane
\be \l{26}
H(z)=\int_a^b{{\ro (\la)}\o {z-\la}}d\la.
\ee
This function is analytic on the entire complex plane except for a cut on the positive
real axis in the interval $[a,b]$. Then one has
\be\l{27}
H(z\pm i\epsilon )=R(z)\mp i\pi\ro (z) \ \ \ , \ \  \ \ b\le z\le a,
\ee
where $R(z)$, from eq.(\r{21}), is
\bea \l{28}
R(z)&=& z\sum_{n=1}{{na_n}\o {\b^{2n-1}}}(z^2-c)^{n-1}-d+{{d-2}\o 2}
\int_a^b{{\ro (z')}\o {z+z'}}dz'\cr \cr
&=& z\sum_{n=1}{{na_n}\o {\b^{2n-1}}}(z^2-c)^{n-1}-d-{{d-2}\o 2}H(-z).
\eea
The constraints (\r{23}) and (\r{25}) for weak and strong--coupling regimes,
respectively, result in $H_w(0)\ge -2$ and $H_s(0)=-2$. But, from
eqs. (\r{27}) and (\r{28}), it can be shown that $H(0)$ must satisfy
$H(0)=-d-{{d-2}\o 2}H(0)
-i\pi\ro (0)$, or $H(0)=-2-(2\pi i/d)\ro (0)$. So the weak density function must
satisfy
\be\l{29}
-i\ro_w(0)\ge 0,
\ee
and strong density function must satisfy
\be\l{30}
-i\ro_s(0)= 0.
\ee
Therefore, the analytical properties of $H(z)$ in the two regimes must be such that
the constraints (\r{29}) and (\r{30}) are satisfied. Using the standard method of
solving the integral equations \c{dc}, one can show that the expression
\be\l{31}
H_w(z)={1\o {2\pi i}}\sqrt{(z-a)(z-b)}\oint_c{{R_w(\la )d\la}\o {(z-\la )
\sqrt{(\la-a)(\la-b)}}},
\ee
has the correct analytical behaviour in weak--coupling region. The contour $c$
in (\r{31}) is a contour encircling the cut $[a,b]$ and excluding $z$. Deforming
$c$ to a contour around the point $z$ and the contour $c_{\infty}$ (a contour
at the infinity), one finds
\be\l{32}
H_w(z)=R_w(z)+{1\o {2\pi i}}\sqrt{(z-a)(z-b)}\oint_{c_{\infty}}{{R_w(\la )
d\la}\o {(z-\la ) \sqrt{(\la-a)(\la-b)}}}.
\ee
Remembering $c_w=0$ and using eqs.(\r{27}) and (\r{28}), it can be easily shown
that the saddle--point equation (\r{21}) reduces to the following expression for
the density function $\ro_w(z)$
\bea\l{33}
\ro_w(z)={{\sqrt{(b-z)(z-a)}}\o \pi}&\{&\sum_{n,p,q}{{na_n}\o {\b^{2n-1}}}
C_pC_{2n-p-q-2}z^qa^pb^{2n-p-q-2} \cr
&&-{{d-2}\o 2}\int_a^b{{dy}\o {y+z}}{{\ro_w(y)}  \o {\sqrt{(b+y)(y+a)}}}\}
\ \ \  \ \ \ {\rm for} \ \ \b>\b_c,
\eea
where
\be\l{34}
C_m={{(2m-1)!!}\o {2^mm!}}.
\ee
In the strong--coupling regime, one can again show that
\bea\l{35}
H_s(z)&=&
{z\o {2\pi i}}\sqrt{{{z-b}\o {z-a}}}\oint_c{{d\la }\o \la}
\sqrt{{{\la-a}\o {\la-b}}}{{R_s(\la )}\o {z-\la }} \cr\cr
&=&R_s(z)+
{z\o {2\pi i}}\sqrt{{{z-b}\o {z-a}}}\oint_{c_{\infty}}{{d\la }\o \la}
\sqrt{{{\la-a}\o {\la-b}}}{{R_s(\la )}\o {z-\la }}
\eea
and, as $c_s=a^2$ (eq.(\r{24})), arrive at
\bea\l{36}
\ro_s(z)=-{z\o \pi}\sqrt{{{b-z}\o {z-a}}}&\{& \sum_{n,m,p,q}
{{na_n}\o {\b^{2n-1}}}(-a^2)^p {{n-}1 \choose p}B_mC_{2n-2p-m-q-2}z^qa^m
b^{2n-2p-m-q-2}  \cr \cr
&&+ {{d-2}\o 2}\int_a^b{{dy}\o {y+z}}
\sqrt{{{y+a}\o {y+b}}}{{\ro_s(y)}\o y} \} \ \ \  \ \ \ {\rm for} \ \ \b<\b_c,
\eea
in which
\be\l{37}
B_m={{(2m-3)!!}\o {2^mm!}}.
\ee
with $B_0\equiv -1$. It can be seen that in the special case $a_n=\delta_{n,1}$
(which corresponds to $V(AA^{\d})=$Tr$(AA^{\d})$), the eqs.(\r{33}) and
(\r{36}) reduce to the corresponding equations in \c{rc3}.

As it is obvious from eqs.(\r{33}) and (\r{36}), here the situation is much more
involved than in SCM, and the most of analytical calculations
done in \c{rc3} can not be done here. One of these kinds of
calculations is discussed in the next section. Therefore, let us restrict
ourselves to the case
$d=2$, for which from eqs.(\r{33}) and (\r{36}) the density functions are known.

Other quantities that must be determined in $d=2$ are the values of parameters
$a$ and $b$ in the both regimes. To find these parameters, we note that at
$|z|\rightarrow \infty$, eqs.(\r{26}) and (\r{22}) imply that $H(z)\rightarrow
1/z$ or $\left( 1/\sqrt{(z-a)(z-b)} \ \ \right) H(z)\rightarrow 1/z^2$. Therefore,
we can expand $\left(1/\sqrt{(z-a)(z-b)}\ \ \right) (R_w(z)-i\pi\ro_w (z))$
(at $d=2$), and take the coefficients
of $1/z$ and $1/z^2$ equal to $0$ and $1$, respectively. In this way we find
the following equations which must be solved to determine $a$ and $b$ in
$\b >\b_c$
\be\l{38}
\sum_{n,m}{{na_n}\o {\b^{2n-1}}}C_mC_{2n-m-1}a^mb^{2n-m-1}-2=0,
\ee
and
\be\l{39}
\sum_{n,m}{{na_n}\o {\b^{2n-1}}}C_mC_{2n-m}a^mb^{2n-m}-(a+b)=1.
\ee
In the strong region, as $(1/z)\sqrt{(z-a)/(z-b)}H(z)
\rightarrow 1/z^2$ at $|z|\rightarrow \infty$, we can again expand
$(1/z)\sqrt{(z-a)/(z-b)}(R_s(z)-i\pi\ro_s (z))$ (at $d=2$) and take the
coefficients
of $1/z$ and $1/z^2$ equal to $0$ and $1$, respectively, and then find the
following equations which determine $a$ and $b$ in $\b <\b_c$
\be\l{40}
\sum_{n,m,p}{{na_n}\o {\b^{2n-1}}}(-a^2)^p{{n-1} \choose p}B_mC_{2n-2p-m-1}a^m
b^{2n-2p-m-1}+2=0,
\ee
and
\be\l{41}
-\sum_{n,m,p}{{na_n}\o {\b^{2n-1}}}(-a^2)^p{{n-1} \choose p}B_mC_{2n-2p-m}a^m
b^{2n-2p-m}+a-b=1.
\ee
Both sets of equations (\r{38}), (\r{39}) and (\r{40}), (\r{41}) are
too complicated to be solved exactly. We discuss about the solutions of these
equations in section 4.

\section{ Some general results for $V=$Tr$(AA^{\d})^n$ }

Let us first derive an expression for the internal energy in terms of $\ro (z)$
for arbitrary $d$. If we denote the internal energy per unit link by $U_{d,V}$,
then the first two equations of (\r{2}) lead to
\be\l{42}
{{d(d-1)}\o 2}U_{d,V}={1\o 2}{{\partial F_{d,V}}\o {\partial \b}}={1\o {2N^2}}
{\partial \o {\partial \b}}({\rm ln} \tilde Z_{d,V}-{\rm ln} \tilde Z_{0,V}),
\ee
where $d(d-1)/2$ is the number of links. Note that ln$\tilde Z_{d,V}\approx
-S_{d,V}\{ {\bar x}_i \}$, where ${\bar x}_i$'s are the solutions of the
saddle--point equation (\r{18}) with density functions (\r{33}) and (\r{36})
in the weak and strong regions, respectively, and $S_{d,V}\{ x_i \}$ is given
by eq.(\r{17}). Therefore, using eq.(\r{17}) for $V=\sum_i(x_i/\b^2)^n$, the
eq.(\r{42}) reduces to
\be\l{43}
d(d-1)U_{d,n}={{2n-1}\o {N\b^{2n}}}\sum_i{\bar x}_i^n-d-{1\o {N^2}}
{\partial \o {\partial \b}}{\rm ln} \tilde Z_{0,n}.
\ee
Now it is interesting that one can calculate the last term of
the above equation exactly. In $d=0$, eq.(\r{11}) is
\bea\l{44}
\tilde Z_{0,n}&=&\int_0^{\infty} \prod_{i=1}^Ndx_i\prod_{i>j}(x_i-x_j)^2e^{-N\b
\sum_{i=1}^N(x_i/\b^2)^n}\cr
&=&\sum^N_{\a_1,\cdots ,\a_N=0\atop {\sum_i\a_i=N(N-1)}}
D_{\a_1,\cdots ,\a_N}\int dx_1\cdots dx_Nx_1^{\a_1}\cdots x_N^{\a_N}
e^{-N\b \sum_{i=1}^N(x_i/\b^2)^n} \cr
&=&\sum^N_{\a_1,\cdots ,\a_N=0\atop {\sum_i\a_i=N(N-1)}}
D_{\a_1,\cdots ,\a_N}I_{\a_1} \cdots I_{\a_N},
\eea
where $D_{\a_1,\cdots ,\a_N}$ are some unimportant constants and $I_{\a}$ is
\bea\l{45}
I_{\a}&=& \int_0^{\infty}dxx^{\a}e^{-N\b (x/\b^2)^n} \cr
&=&{1\o n}\left( {{\b^{2n-1}}\o N}\right)^{(\a +1)/n}\Gamma ({{\a +1}\o n}).
\eea
Therefore
\be\l{46}
\tilde Z_{0,n}=\b^{N^2(2n-1)/n}\times (\b {\rm -independent \ \ terms}),
\ee
and from this, eq.(\r{43}) becomes
\be\l{47}
d(d-1)U_{d,n}={{2n-1}\o {N\b^{2n}}}\sum_i{\bar x}_i^n-d+({1\o n}-2){1\o \b}.
\ee
In the large--$N$ limit, this equation for weak and strong regimes becomes
\be\l{48}
d(d-1)U^{(w)}_{d,n}={{2n-1}\o {\b^{2n}}}\int_a^bz^{2n}\ro_w(z)dz-d
+({1\o n}-2){1\o \b},
\ee
and
\be\l{49}
d(d-1)U^{(s)}_{d,n}={{2n-1}\o {\b^{2n}}}\int_a^b(z^2-a^2)^n\ro_s(z)dz-d
+({1\o n}-2){1\o \b},
\ee
respectively. Again, in the case $n=1$, where the model is SCM, the above
equations reduce to the corresponding equations found in \c{rc3}. As can
be seen from these relations, the power of variables in the integrands makes the
integrations difficult
to perform, even when the density functions are known. Let us show
this by an example. It can be shown that in $d>4$, the critical value of $a$ is
different from zero ($a_c\neq 0$) \c{rc3}, and of course at this point
$\ro_s=\ro_w=\ro_c$.
Therefore, at the critical point we have
\be\l{50}
d(d-1)[U^{(s)}_{d,n}-U^{(w)}_{d,n}]={{2n-1}\o {\b_c^{2n}}}\int_{a_c}^{b_c}
[(z^2-a_c^2)^n-z^{2n}]\ro_c(z)dz.
\ee
For SCM ,where $n=1$, the above difference reduces to (using eq.(\r{22}))
\be\l{51}
d(d-1)[U^{(s)}_{d,1}-U^{(w)}_{d,1}]=-{{a_c^2}\o {\b^2_c}}\int_{a_c}^{b_c}
\ro_c(z)dz=-{{a_c^2}\o {\b^2_c}}\neq 0,
\ee
and this simply proves that there exists a first--order phase transition in SCM
with $d>4$, although we do not know the explicit form of $\ro_c(z)$. But if $n$
is different from $1$, the same calculation is not possible. For example if
$n=2$, eq.(\r{50}) gives
\be\l{52}
d(d-1)[U^{(s)}_{d,2}-U^{(w)}_{d,2}]={{3a_c^2}\o {\b^4_c}}[a_c^2-2\int_{a_c}^{b_c}
\ro_c(z)z^2dz].
\ee
In this case one must know $\ro_c(z)$ to say something about the result.

So let us again restrict ourselves to $d=2$, and try to find the critical value
$\b_c$ for $V=$Tr$(AA^{\d})^n$. As mentioned above, at the critical point
$\ro_s(z)=\ro_w(z)=\ro_c(z)$ and therefore the constraint (\r{30}) must also
hold for $\ro_w(z)$ at this point,{\it i.e.}, $\ro_w^{{\rm critical}}(0)=0$. Using
(\r{33}) for $d=2$ and putting $z=0$, we obtain
\be\l{53}
\sqrt{a_cb_c}\sum_{p=0}C_pC_{2n-2p-2}a_c^pb_c^{2n-2p-2}=0.
\ee
As the coefficients of all terms in the above equation are positive, and
$0\leq a_c<b_c$, the only nontrivial solution of (\r{53}) is
\be\l{54}
a_c=0 \ \ \  , \ \ \ {\rm for}\ \ d=2.
\ee
Therefore, at the critical point eqs.(\r{40}) and (\r{41}) reduce to
\be\l{55}
{n\o {\b_c^{2n-1}}}B_0C_{2n-1}b_c^{2n-1}+2=0,
\ee
and
\be\l{56}
-{n\o {\b_c^{2n-1}}}B_0C_{2n}b_c^{2n}-b_c=1,
\ee
respectively. From these two equations, one can find the critical values $b_c$
and $\b_c$ for $V=$Tr$(AA^{\d})^n$ in $d=2$ as following
$$ b_c={{2n}\o {2n-1}}, $$
\be\l{57}
\b_c={{2n}\o {2n-1}}\left[ {{n(4n-3)!!}\o {2^{2n}(2n-1)!}}\right]^{{1\o {2n-1}}}.
\ee
In the case $n=1$ (SCM), eq.(\r{57}) gives $\b_c=1/2$, which is the correct value of
$\b_c$ in the Gross--Witten single--link problem.

\section { Phase structure in d=2}

In this section we study the phase structure of some special cases of gSCM in
$d=2$ with $V=$Tr$(AA^{\d})^n$.

\subsection { $n=1$ (YM$_2$)}

In the weak--coupling regime, eqs.(\r{38}) and (\r{39}) are linear equations with
solutions $a=2\b -\sqrt{2\b}$ and $b=2\b +\sqrt{2\b}$, and therefore eq.(\r{33})
leads to
\be\l{58}
\ro_w^{(n=1)}(z)={{\sqrt{(b-z)(z-a)}}\o {\pi\b}}=
{{\sqrt{2\b-(z-2\b )^2}}\o {\pi\b}}.
\ee
In the strong--coupling regime, eqs.(\r{40}) and (\r{41}) result $a=1/2-\b$ and
$b=1/2+\b$, and therefore eq.(\r{36}) leads to
\be\l{59}
\ro_s^{(n=1)}(z)={z\o \pi\b}\sqrt{{{b-z}\o {z-a}}}
={z\o \pi\b}\sqrt{{{1+6\b -2z}\o {2z-1+2\b}}}.
\ee
These equations are the density functions that have been obtained in \c{rc3}
(note that in this paper, $x_i$'s are the eigenvalues of $\b^2AA^{\d}$, but
in ref.\c{rc3} $x_i$'s are the eigenvalues of $4\b^2AA^{\d}$). If one computes
the internal energies (\r{48}) and (\r{49}) by using these density functions,
it is seen that
\be\l{60}
U_w^{(n=1)}=1-{1\o {4\b}}  \ \ \  , \ \ \ {\rm for}\ \ \b\geq {1\o 2},
\ee
and
\be\l{61}
U_s^{(n=1)}=\b  \ \ \  , \ \ \ {\rm for}\ \ \b\leq {1\o 2},
\ee
which are the results obtained in \c{dg}. It is obvious from
eqs.(\r{60}) and (\r{61}) that in the case $n=1$, there exists a third--order phase
transition.

\subsection { $n=2$ gSCM}

In this case the eq.(\r{57}) leads to
$$b_c^{(n=2)}={4\o 3}, $$
\be \l{62}
\b_c^{(n=2)}={2\o 3}\left( {5\o 2} \right)^{1/3}.
\ee
First consider the weak--coupling regime. The eqs.(\r{38}) and (\r{39})
give, respectively
\be\l{63}
5(a^3+b^3)+3ab(a+b)=16\b^3,
\ee
and
\be\l{64}
35(a^4+b^4)+20ab(a^2+b^2)+18a^2b^2=64\b^3(a+b+1).
\ee
These equations can not be solved analytically to obtain $a$ and $b$, but as we want
to study the phase structure of the model, it is sufficient to look at the
solutions near the critical point. Therefore we expand the above equations around
$a_c=0$ and $b_c=4/3$ up to second order, and then find $a$ and $b$ in terms of
$\a =\b -\b_c$, again up to second order. The final result is
$$
a_w={1\o 4}5^{2/3}2^{1/3}\a -{3\o 64}5^{1/3}2^{2/3}\a^2+\cdots ,
$$
\be\l{65}
b_w={4\o 3}+{7\o 20}5^{2/3}2^{1/3}\a -{9\o 320}5^{1/3}2^{2/3}\a^2+\cdots .
\ee
Now consider the density function $\ro (z)$ in the weak--coupling regime
for $a_n=\delta_{n,2}$ and $d=2$ (eq.(\r{33})):
\be\l{66}
\ro_w^{(n=2)}(z)={2\o {\pi\b^3}}\left[ {3\o 8}(a^2+b^2)+{1\o 4}ab+{1\o 2}(a+b)z
+z^2\right] \sqrt{(b-z)(z-a)},
\ee
using this function, we calculate (using (\r{48})) the internal energy
(per unit link) in the weak--coupling regime
\bea\l{67}
U_w^{(n=2)}&=&{3\o {2\b^4}}\int_a^b\ro_w(z)z^4dz-1-{3\o 4\b } \cr
&=&{9\o {2^{12}\b^7}}\left[ {315\o 8}(a^8+b^8)-7ab(a^6+b^6)-{19\o 2}a^2b^2
(a^4+b^4) -17a^3b^3(a^2+b^2)-{47\o 4}a^4b^4\right]\cr && -1-{3\o {4\b}}.
\eea
If we substitute the expansions (\r{65}) in the above relation, we find the
internal energy in terms of $\b$, up to order $\a^2$, as following
\be\l{68}
U_w^{(n=2)}={279\o 400}5^{2/3}2^{1/3}+1 -{333\o 800}5^{1/3}2^{1/3}\a
+{189\o 160}\a^2+\cdots .
\ee
We can follow the same steps in the strong--coupling regime, this time
using eqs.(\r{40}), (\r{41}), (\r{36}), and (\r{49}). The final results are
$$
a_s=-{1\o 4}5^{2/3}2^{1/3}\a -{3\o 16}5^{1/3}2^{2/3}\a^2+\cdots ,
$$
$$
b_s={4\o 3}+{7\o 20}5^{2/3}2^{1/3}\a +{9\o 80}5^{1/3}2^{1/3}\a^2+\cdots ,
$$
\be\l{69}
U_s^{(n=2)}={279\o 400}5^{2/3}2^{1/3}+1 -{333\o 800}5^{1/3}2^{1/3}\a
+{81\o 40}\a^2+\cdots ,
\ee
and therefore
\be\l{70}
U_s^{(n=2)}-U_w^{(n=2)}={27\o 32}(\b -\b_c)^2+\cdots .
\ee
The above equation shows that we have a third--order phase transition for
$V=$Tr$(AA^{\d})^2$ of gSCM in $d=2$.

\subsection { $n=3$ and $n=4$ gSCM}

In the case $n=3$, $b_c=6/5$ and $\b_c=(3/80)(16)^{4/5}(189)^{1/5}$;
and in the case $n=4$, $b_c=8/7$ and $\b_c=(1/14)3^{1/7}8^{6/7}(143)^{1/7}$.
The procedure is the same as $n=2$ and after some calculations, we obtain
the final results as following
\be\l{71}
U_s^{(n=3)}-U_w^{(n=3)}={1525\o 23814}(378)^{2/5}(\b -\b_c)^2+\cdots ,
\ee
\be\l{72}
U_s^{(n=4)}-U_w^{(n=4)}={343\o 16896}(429)^{4/7}2^{2/7}(\b -\b_c)^2+\cdots .
\ee
These relations show that for $V=$Tr$(AA^{\d})^3$ and Tr$(AA^{\d})^4$ of gSCM
in $d=2$, the models exhibit third--order phase transition, as in the YM
theory. We expect that the same behaviour exists in all
gSCM with $V=$Tr$(AA^{\d})^n$.

\section { Conclusion}

In this paper the SCM was generalized by a method very similar to the
method which have been used in the generalization of YM theories.
This generalization may be interesting from several points of view. First, as
mentioned in the introduction, many few  matrix--models can be solved
analytically in the large--$N$ limit. gSCM's are models that
can be studied analytically in this limit. Second, the gSCM at $d=2$ can be
treated as the lattice version of gYM theory, which is an important two--dimensional
candidate of QCD. And third, which is somehow related to the second, the study
of these models (gSCM) is much more simple than the gYM. For instance, the proof
of the existence of third--order phase transition in $f(B)=$Tr$B^4$ gYM theory
\c{ma1}, which corresponds to our $V=$Tr$(AA^{\d})^2$ model, is more difficult
than the proof which is given in section 4.2.

\vspace {1cm}

\noindent{\bf Acknowledgement}

\noindent I would like to thank the research council of the
University of Tehran, for partial financial support.


\vskip 1cm


\begin{thebibliography}{99}
\bibitem{kw} K. Wilson; Phys. Rev. {\bf D10} (1974) 2445.
\bibitem{he} H. E. Stanley; Phys. Rev. {\bf 176} (1968) 718.
\bibitem{gt} G. 't Hooft; Nucl. Phys. {\bf B72} (1974) 461.
\bibitem{dg} D. J. Gross and E. Witten; Phys. Rev. {\bf D21} (1980) 446.
\bibitem{rc1} R. C. Brower and M. Nauenberg; Nucl. Phys. {\bf B180} (1981) 221.
\bibitem{eb} E. Brezin and D. J. Gross; Phys. Lett. {\bf B97} (1980) 120.
\bibitem{rc2} R. C. Brower, P. Rossi and C. -I. Tan; Phys. Rev. {\bf D23}
(1981) 942; {\bf D23} (1981) 953.
\bibitem{df} D. Friedan; Commun. Math. Phys. {\bf 78} (1981) 353.
\bibitem{pr1} P. Rossi, M. Campostrini and E. Vicari; Phys. Rep. {\bf 302} (1998) 143.
\bibitem{pr2} P. Rossi and C. -I. Tan; Phys. Rev. {\bf D51} (1995) 7159.
\bibitem{rc3} R. C. Brower, M. Campostrini, K. Orginos, P. Rossi, C. -I. Tan,
and E. Vicari; Phys. Rev. {\bf D53} (1996) 3230.
\bibitem{ew} E. Witten; Jour. Geom. Phys. {\bf 9} (1992) 303.
\bibitem{mr} M. R. Douglas, K. Li and M. Staudacher; Nucl. Phys. {\bf B240}
             (1994) 140;\\
             O. Ganor, J. Sonnenschein, and S. Yankielowicz;
             Nucl. Phys. {\bf B434} (1995) 139.
\bibitem{mk} M. Khorrami and M. Alimohammadi; Mod. Phys. Lett. {\bf A12} (1997)
             2265.
\bibitem{br} B. Rusakov and S.Yankielowicz; Phys. Lett. {\bf B339} (1994)
            258.
\bibitem{ma1} M. Alimohammadi, M. Khorrami and, A. Aghamohammadi;
             Nucl. Phys. {\bf B510} (1998) 313.
\bibitem{ma2} M. Alimohammadi and A. Tofighi;
 Eur. Phys. Jour. {\bf C8} (1999) 711.
\bibitem{tr} T. R. Morris; Nucl. Phys. {\bf B356} (1991) 703.
\bibitem{dc} D. Gakhov; ``Boundary problems'', Russian edition
             (Nauka, 1975),\\
              A. C. Pipkin; ``A course on integral equations''
             (Springer, Berlin, 1991).

\end{thebibliography}
\end{document}